# Wireless power transfer via topological modes in dimer chains

Juan Song[1], Fengqing Yang[1], Zhiwei Guo[1,*], Xian Wu[1], Kejia Zhu[2], Jun Jiang[3], Yong Sun[1], Yunhui Li[1,†], Haitao Jiang[1], and Hong Chen[1]

[1]MOE Key Laboratory of Advanced Micro-structured Materials, School of Physics Sciences and Engineering, Tongji University, Shanghai 200092, China

[2]Department of Electrical Engineering, Tongji University, Shanghai 201804, China

[3]School of Automotive Studies, Tongji University, Shanghai 210804, China

**Abstract**

The topological characteristics, including invariant topological orders, band inversion, and the topological edge mode (TEM) in the photonic insulators, have been widely studied. Whether people can take advantage of intriguing topological modes in simple one-dimensional systems to implement some practical applications is an issue which people are increasingly concerned about. In this work, based on a photonic dimer chain composed of ultra-subwavelength resonators, we verify experimentally that the TEM in the effective second-order parity-time (*PT*) system is immune to the inner disorder perturbation, and can be used to realize the long-range wireless power transfer (WPT) with high transmission efficiency. To intuitively show the TEM can be used for WPT, a power signal source is used to excite the TEM. It can be clearly seen that two LED lamps with 0.5-W at both ends of the structure are lighted up with the aid of TEMs. In addition, in order to solve the special technical problems of standby power loss and frequency tracking, we further propose that a WPT system with effective third-order *PT* symmetry can be constructed by using one topological interface mode and two TEMs. Inspired by the long-range WPT with TEMs in this work, it is expected to use more complex topological structures to achieve energy transmission with more functions, such as the WPT devices whose direction can be selected flexibly in the quasiperiodic or trimer topological chains.

**Keywords:** Topological edge modes; Su-Schrieffer-Heeger model; Wireless power transfer.

J. Song and F. Q. Yang contribute equally to this work.

* 2014guozhiwei@tongji.edu.cn, † liyunhui@tongji.edu.cn

# I. INTRODUCTION

Using topology to control the motion of photons is of great scientific significance. This unique research has developed into a new research topic, namely topological photonics [1-3]. Photonic topological edge modes (TEMs) can overcome the scattering losses caused by structural defects and disorders, which have been proposed for numerous topologically protected photonic devices, such as unidirectional waveguides [4-8], robust delay lines [9-12] and single-mode lasers [13-16]. The photonic dimer chain is mapped from the basic topological Su-Schrieffer-Heeger (SSH) model in condensed matter physics [17]. Notably, in 2009 Malkova *et al* experimentally revealed the linear Shockley-like surface states in an optically induced semi-infinite photonic superlattice and it is the first demonstration of one-dimensional topological states in photonics [18]. In the early research of metamaterials, the coupling process between split-ring resonators was described. The change of magnetic resonance coupling intensity with distance was also reported [19]. Specially, the topological phase transition, topological orders and the associated TEMs in the topological dimer chain have been experimentally observed in the platform composed of split-ring resonators [20]. The great progress of the topological dimer chains has enabled some interesting extensions to be realized associated with some important applications [21, 22]. Recently, research on the advantages of dimer chains has been extended to nonlinear [23-25] and active [26, 27] systems, where the active-controlled TEMs and topological lasers are studied, respectively. Moreover, the topological properties of the non-Hermitian dimer chain have been theoretically analyzed and experimentally demonstrated [28-32]. Especially, combining the non-Hermitian exceptional points (EPs) with topological modes, a new sensor which is not sensitive to internal manufacturing errors but is highly sensitive to changes in the external environment is proposed [33]. Therefore, although photonic dimer chain structure is simple, its fertile topological physics and useful applications have attracted more and more attention.

At present, wireless power transfer (WPT) has triggered immense research interest in a range of practical applications, including mobile phones, robots, medical-implanted devices and electric vehicles [34-38]. The traditional WPT devices based on magnetic induction mechanism are usually seriously limited by the transmission distance [39]. When the distance between the receiving coil and the transmitting coil is large, the transmission efficiency will be greatly reduced. Recently, although a variety of metamaterials have been demonstrated to be powerful tools to improve the functionalities and obtain new performance of WPT systems, the array structures are complex [40-43]. As another effective solution, magnetic resonance WPT is proposed to achieve mid-range efficient energy transmission [44-54]. However, the corresponding working frequency needs to be tuned with the transmission distance because of the near-field coupling mechanism, which limits its practical applications. To solve the problem of frequency tracking, optimization schemes are proposed, for example, use nonlinear effect [55-57] and active feedback circuit [58] to track the real-time working frequency or resort to the novel high-order *PT* system [59-62]. So far, the proposed WPT devices usually have inherent sensitivity to transmission distance and structural disturbance, which is an essential challenge for various application scenarios, such as the robots whose structure will often deform. A natural question arises: can we design new WPT devices with topological properties?

To solve this problem, we recently proposed that the nontrivial dimer chain will provide a suitable platform for people to study the robust WPT in RF regime [63]. On the one hand, similar to the Domino structure composed of the coupled resonators for long-range WPT [64], the topological dimer chain can be used to realize the efficient long-range WPT. On the other hand, the TEMs in nontrivial dimer chain are topological protected, thus the corresponding WPT will robust against the disorders and fluctuations. In this work, by using the ultra-subwavelength coil resonators, we design and fabricate the dimer chains under the tight-binding model and apply the TEMs of topological chain to the long-range WPT. The TEM of nontrivial dimer chain offers great practical advantages in the technology and applications of WPT, such as long-range transfer and robustness. However, there are some aspects of the TEMs in dimer chain for long-range WPT that should be noted. (1) The TEMs at the two ends of a finite topological dimer chain will split as a result of near-field coupling. Therefore, the working frequency needs to be tuned according to different load power or chain length. In particular, only when the system reaches EP, TEMs splitting induced by near-field coupling can be eliminated and accompanied by efficient transmission. (2) The standby power loss of the topological dimer chain is obvious, which not only damages the WPT devices, but also wastes unnecessary energy. In order to overcome the above mentioned two problems, we further propose an effective third-order *PT* symmetry WPT system in a composite dimer chain. As we have previously studied based on three simple coils, high-order *PT* symmetry can be used to achieve WPT without frequency tracking and remain low standby power loss [60]. This paper uses a topological interface mode (TIM) and two TEMs in the composite topological dimer chain to realize the effective third-order *PT* symmetric and then realize the high-performance long-range WPT. Our results provide a versatile platform to design the long-range WPT devices with topological protection. In addition, considering the asymmetric TEMs in the quasiperiodic [65] or the trimer [66] topological chains, our platform is also expected to further realize the interesting and directional WPT.

This paper is organized as follows: Sec. II uncovers the design of the topological dimer chain and the demonstration of the robust TEMs with topological protection; in Sec. III, the physical mechanism of the effective second-order *PT* system with TEMs in the dimer chain is analyzed, and both calculations and experiments are carried out to verify the long-range WPT; in Sec. IV, an effective third-order *PT* system with two TEMs and one TIM is studied, and the problems of standby power loss and frequency tracking can be well solved. Finally, Sec. V summarizes the conclusions of this work.

## II. CHARACTERIZATION OF TEMS IN DIMER CHAINS

The topological dimer chains are constructed by ultra-subwavelength coil resonators. Specially, the composite coil resonator is composed of two layers divided by a polymethyl methacrylate (PMMA) substrate with a thickness of $h = 1$ cm. Figure 1(a) shows the dimer chain with $N_c$ unit cells. The top layer corresponds to a coil resonator and the photo is shown in Fig. 1(b). All the coil resonators in the dimer chain are the same. The corresponding resonant frequency is 5.62 MHz, which depends on the loaded lumped capacitor $C$ = 100 pF and the geometric parameters, including the inner diameter $D_1$ = 5.2 cm and outer diameter $D_2$ = 7 cm. Specially, for the spiral ring resonator loaded with lumped capacitance, the electrical resonance coupling between the resonators can be ignored. And the type of coupling here is magnetic resonance coupling. Figure 1(c) shows the back layer of the composite resonator, in which a 0.5-W LED lamp is connected to a non-resonant coil. Once the magnetic field in the top coil is strong enough, the LED lamp loaded in the back layer can be lighted up. Based on the tight-binding model, the Hamiltonian of the photonic dimer chain can be written as follows [17]:

$$H = \sum_n (\kappa_1 a_n^\dagger b_n + \kappa_2 a_{n+1} b_n^\dagger) + h.c., \qquad (1)$$

where $\kappa_1$ and $\kappa_2$ respectively denote the intra-cell coupling and inter-cell coupling coefficients, which are controlled by the distance between coil resonators [67]. The coupling strength decreases with the increase of coupling distance, owing to the exponentially decaying property of near fields. From the spectral response, we can fit a decaying exponential function of $\kappa$ as $\kappa = 0.98 e^{-d/4.42}$ [67]. In our photonic dimer chain, the distance of intra-cell and inter-cell is $d_1$ = 4 cm and $d_2$ = 3 cm, and the corresponding coupling coefficients are $\kappa_1$ = 0.4 MHz and $\kappa_2$ = 0.5 MHz, respectively. $a_n (b_n)$ and $a_n^\dagger (b_n^\dagger)$ represent the generation and annihilation operators corresponding to $a$ ($b$) lattice point in the $n^{\text{th}}$ unit cell of SSH chain, respectively. In this model, the topological order is related to the relative magnitude of intra-cell and inter-cell coupling coefficients, and the TEMs will appear symmetrically at two ends of the chain for the nontrivial structure ($\kappa_1 < \kappa_2$).

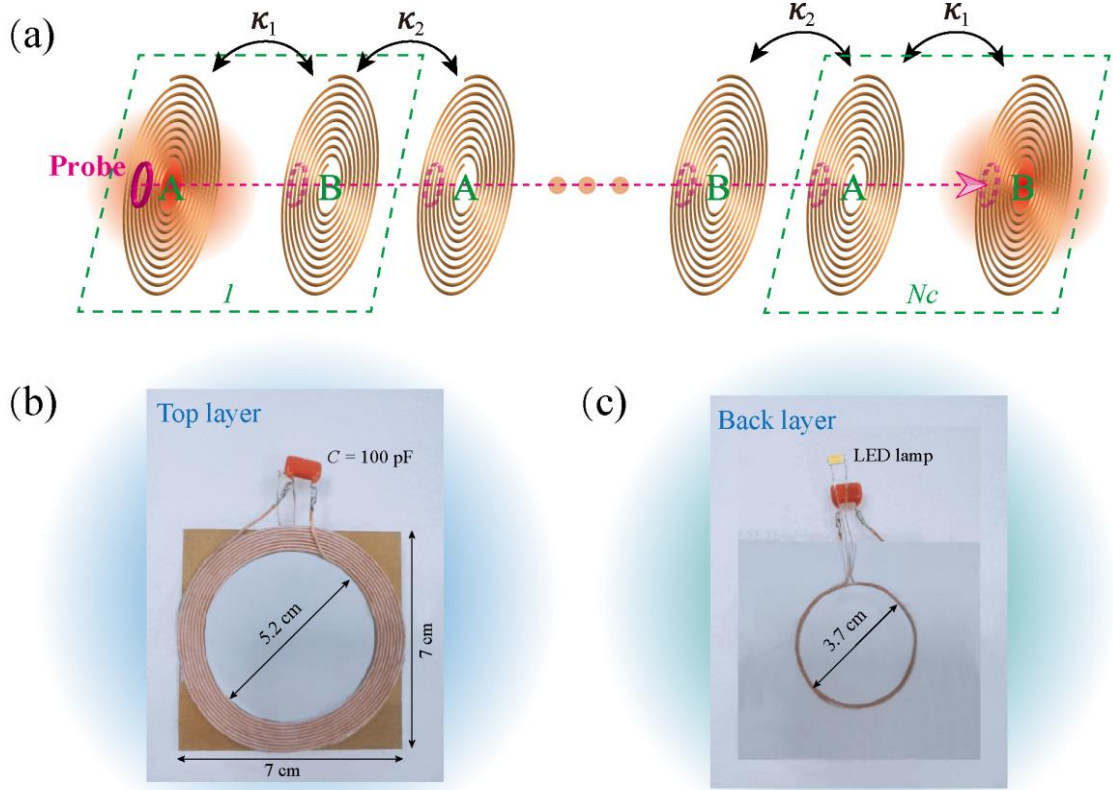

FIG. 1. **Schematic of the topological dimer chain composed of the composite coil resonators**. (a) Schematic of the dimer chain. In this tight-binding model, the unit cell is composed of two resonators, which are marked by a dashed green rhombus to facilitate viewing. (b) The resonant coil on the top layer of the composite structure. Here, the resonant frequency of the coil is 5.62 MHz; the loaded capacitor is 100 pF; the thickness of the substrate is 1 cm; the inner diameter and outer diameter of the coil resonator are 5.2 cm and 7cm, respectively. (c) A non-resonant coil with a LED lamp on the back layer of the composite structure. The corresponding diameter of the non-resonant coil is 3.7 cm.

Based on the method of near-field detection [20, 63], we measure the density of states (DOS) spectrum of the topological dimer chain composed of 16 coil resonators. Our near-field magnetic probe is a loop antenna connected to the port of the vector network analyzer (Agilent PNA Network Analyzer N5222A). The radius of the loop probe is 2 cm. It can be taken as a non-resonant antenna with high impedance. This small loop antenna acts as a source to excite the sample and then measure the reflection, as shown in Fig. 1(a). And the local density of states (LDOS) of each site is obtained from the reflection by putting the probe to the center of the corresponding resonator.

The DOS spectrum is obtained by averaging the LDOS spectral over all 16 sites. The measured DOS spectrum of the topological dimer chain without disorder perturbation is shown in Fig. 2(a). An isolated state (5.62 MHz) exists in the bandgap of the spectrum, which belongs to the TEM shown by red arrow. For comparison, a bulk state (5.43 MHz) near the TEM is selected by green arrow. To investigate the robustness of the TEMs against disorder perturbation, we randomly move 10 coil resonators 5 mm in the center of the topological dimer chain. The corresponding measured DOS spectrum is shown in Fig. 2(b). Compared with Fig. 2 (a) and Fig. 2 (b), it can be found that the TEM is still maintained for the dimer chain with perturbation, whereas the bulk state has been deteriorated seriously.

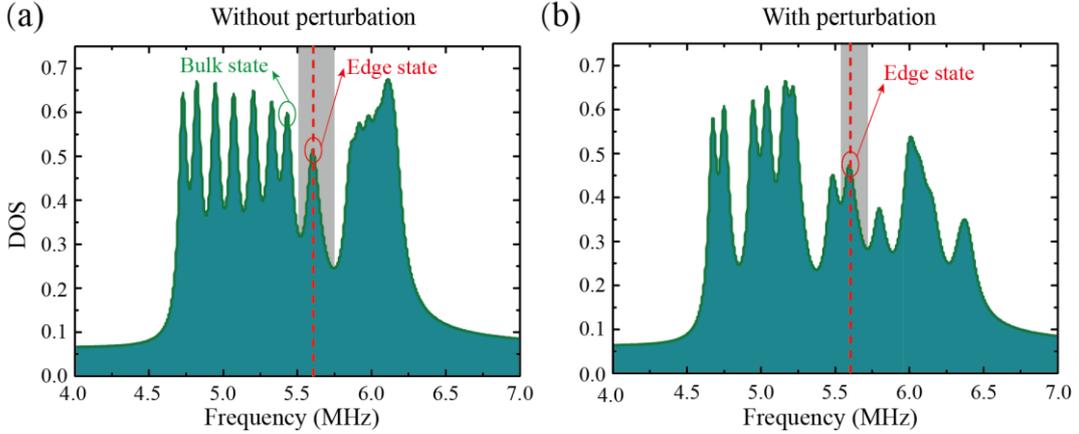

FIG. 2. **Measured DOS spectrum of the topological dimer chains.** (a) Measured DOS spectrum of the topological dimer chain without perturbation. One bulk state (5.43 MHz) and the TEM (5.62 MHz) in the band gap are marked by the green arrow and red arrow, respectively. (b) Similar to (a), but for the case with disorder perturbation in the inner of the chain. The structure disorder is realized by randomly moving the central 10 resonators with 5 mm.

Moreover, we measure the LDOS distributions of the TEM and the bulk state, which are marked by red and green arrows in Fig. 2, respectively. For the bulk state in the topological dimer chain without perturbation, the LDOS is mainly distributed in the bulk, as shown in Fig. 3(a). However, once the topological dimer chain with disorder perturbation, the LDOS of the bulk state is significantly modified in Fig. 3(b). However, for the topological dimer chain with disorder perturbation, the measured LDOS of the TEM is still confined at the two ends of the chain (Fig. 3(d)), just as the case without perturbation (Fig. 3(c)). This indicates that when random movement is introduced into the interior of the chain, the TEM is almost unaffected, which is highly significant for robust WPT. In addition, it should be noted that when the topological dimer chain is long enough, the edge state can be localized at the boundary of the chain, and the wave function decays exponentially with the distance away from the boundary. However, with the shortening of the chain, the localization of topological state will gradually decrease, and the characteristic of exponential decay will also gradually weaken. Therefore, the edge states without ideal local properties in Figs. 3(c) and 3(d) are mainly due to the limited length of the topological dimer chain.

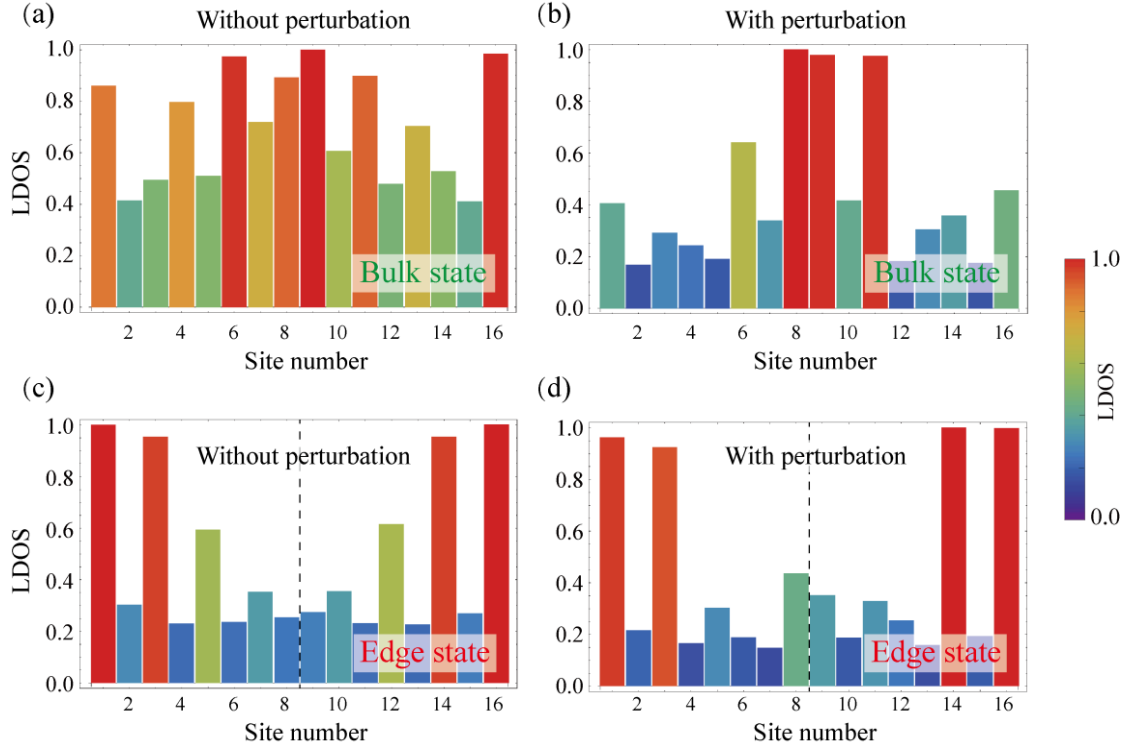

FIG. 3. **Robust TEM against the disorder perturbation.** (a) Measured normalized LDOS distribution of the bulk state (marked by green arrow in Fig. 2(a)) in the topological dimer chain without perturbation. (b) Similar to (a), but for the LDOS distribution of the bulk state with disorder perturbation. (c) (d) Measured normalized LDOS distribution of the TEM in Figs. 2(a) and 2(b), respectively. Colors are used to represent the different values of LDOS. The center of the symmetry is marked with a black dashed line.

### III. LONG-RANGE WPT IMPLEMENTED BY EFFICIENT SECOND-ORDER *PT* SYMMETRIC SYSTEM WITH TEMS

After determining the robust TEMs in the topological dimer chain, we further observe the efficient transmission of TEMs based on the standard multi-coil WPT system, as shown in Fig. 4(a). Two non-resonant coils are placed at the left and right ends of dimer chain as transmitter and receiver, respectively. For a continuous harmonic wave input $s_{1+} = S_{1+} e^{-i\omega t}$, the dynamics of the multi-coil system, which ignoring the intrinsic loss, can be described by the coupled mode equations as [55, 60-62]:

$$\frac{da_1}{dt} = (-i\omega_0 - \gamma)a_1 - i\kappa_1 a_2 + \sqrt{2\gamma}s_{1+}$$

$$\frac{da_n}{dt} = -i\omega_0 a_n - i\kappa_1 a_{n-1} - i\kappa_2 a_{n+1} \quad (n = 2, 4, \ldots 14)$$

$$\frac{da_m}{dt} = -i\omega_0 a_m - i\kappa_2 a_{m-1} - i\kappa_1 a_{m+1} \quad (m = 3, 5, \ldots 15)$$

$$\frac{da_{16}}{dt} = (-i\omega_0 - \gamma)a_{16} - i\kappa_1 a_{15}$$

(2)

where the terms $\gamma = 0.15$ MHz describes the coupling rate between the resonant coil on the end of the dimer chain and the non-resonant coil. $\omega_0 = 5.62$ MHz is the resonant frequency of the resonator coils. $|a_j|^2$ ($j = 1, 2, \ldots 16$) denotes the energy stored in different resonator coils. The power transfer efficiency (transmittance) of the dimer chain can be expressed as $\eta = |S_{21}|^2$, which can be determined by the output wave $s_{2-} = \sqrt{2\gamma}a_{16}$ as $S_{21} = s_{2-}/s_{1+}$. Considering the zero reflected wave $s_{2-} = -s_{1+} + \sqrt{2\gamma_1}a_1 = 0$, the Hamiltonian of the effective second-order open system realized by the two TEMs in the topological dimer chain can be expressed as [60-62]:

$$H_{ee} = \begin{pmatrix} \omega_0 + i\gamma & \kappa_{ee} \\ \kappa_{ee} & \omega_0 - i\gamma \end{pmatrix},$$

(3)

where $\kappa_{ee}$ denotes the effective coupling coefficient of two TEMs. Although the effective Hamiltonian of Eq. (3) is different from that of Eq. (1) of the ideal closed system, the eigenvalues of two TEMs are almost the same. Figures 4(b) shows the calculated transmittance of two topological distinguished dimer chains. The solid blue line and red dashed line denote the results of topological chain and trivial chain, respectively. It is obvious that there is a transmission peak at the position of TEM in the topological dimer chain. The transmission enhancement of topological chain over trivial chain is shown in Fig. 4(c). It can be clearly seen that the transmission enhancement at the working frequency (5.62 MHz) is near 44.63. Therefore, the performance of long-range WPT can be significantly affected by changing the topological configuration of dimer chain. For the topological dimer chain, the associated TEM can be used for long-range WPT. In fact, the maximum effective distance of WPT for TEMs is limited by the loss of the system. With the increase of the number of resonant coils, the transmission distance will be longer, but due to the influence of intrinsic loss and radiation loss of the system, the transmission efficiency will be reduced.

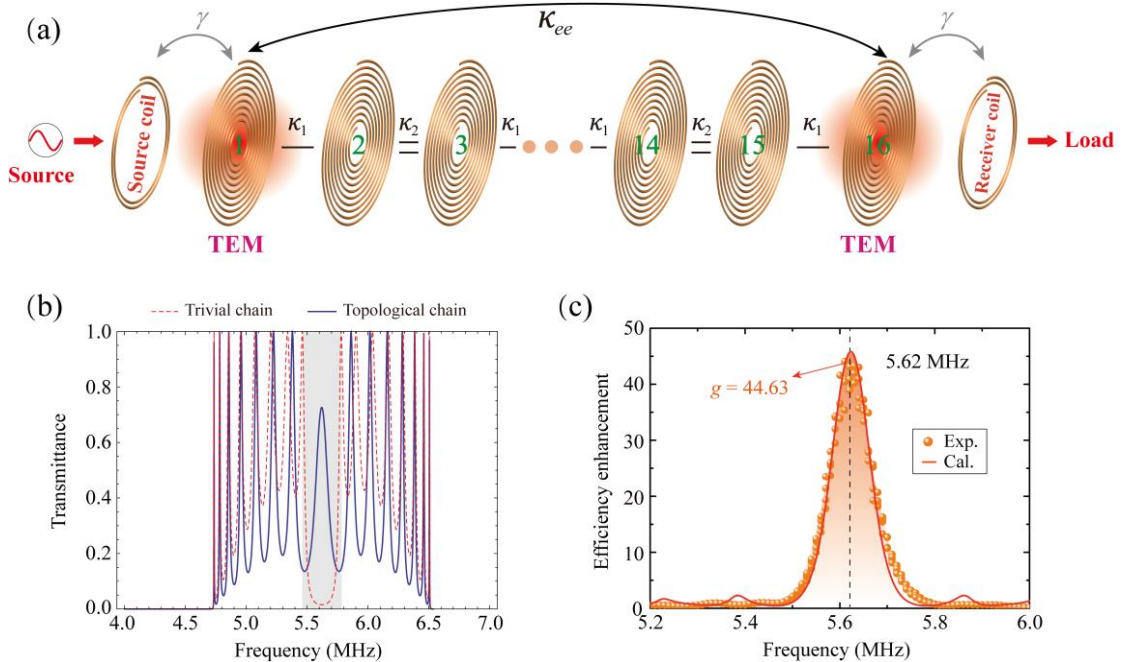

FIG. 4. **Transmission enhancement of topological chain over a trivial chain.** (a) Schematic of a multi-coil WPT system based on the second-order *PT* symmetry of TEMs. The signal is input and output by source coil and receiving coil, respectively. (b) Calculated transmittance spectra as the function of frequency for the topological chain (blue solid line) and trivial chain (red dashed line). (c) The ratio of transmission efficiency of topological chain to that of a trivial chain. At the working frequency 5.62 MHz, the efficiency enhancement is 44.63.

Then, an intuitive experiment is carried out to exhibit the long-range WPT. The high-power signal source (AG Series Amplifier, T&C Power Conversion) instead of the vector network analyzer, is used to excite the TEMs. By placing a source coil to the center of the topological dimer chain, the edge state will be established at both ends of the chain. In such a near-field system, the power transfer is based on the near-filed coupling instead of the far-filed propagating. To show the topological WPT conveniently, except for the two resonators near the source coil, all resonators are added with LED lamps. At the working frequency of TEM, the LED lamps on both sides of the chain are lighted up whereas the LED lamps loaded on other resonators remain dark, as shown in Fig. 5. In fact, the magnetic field is distributed on all resonators near the boundary of the chain. In order to observe the edge state clearly, we tune the signal source to the appropriate power. In this case, the resonators on the boundary can reach the luminous power of the LED lamp, while the magnetic field of other resonators is not enough to light up the LED lamp. Considering the asymmetric field pattern of TEM in complex chains, the unidirectional long-range WPT can also be realized by our setup based on the coil resonators [68, 69].

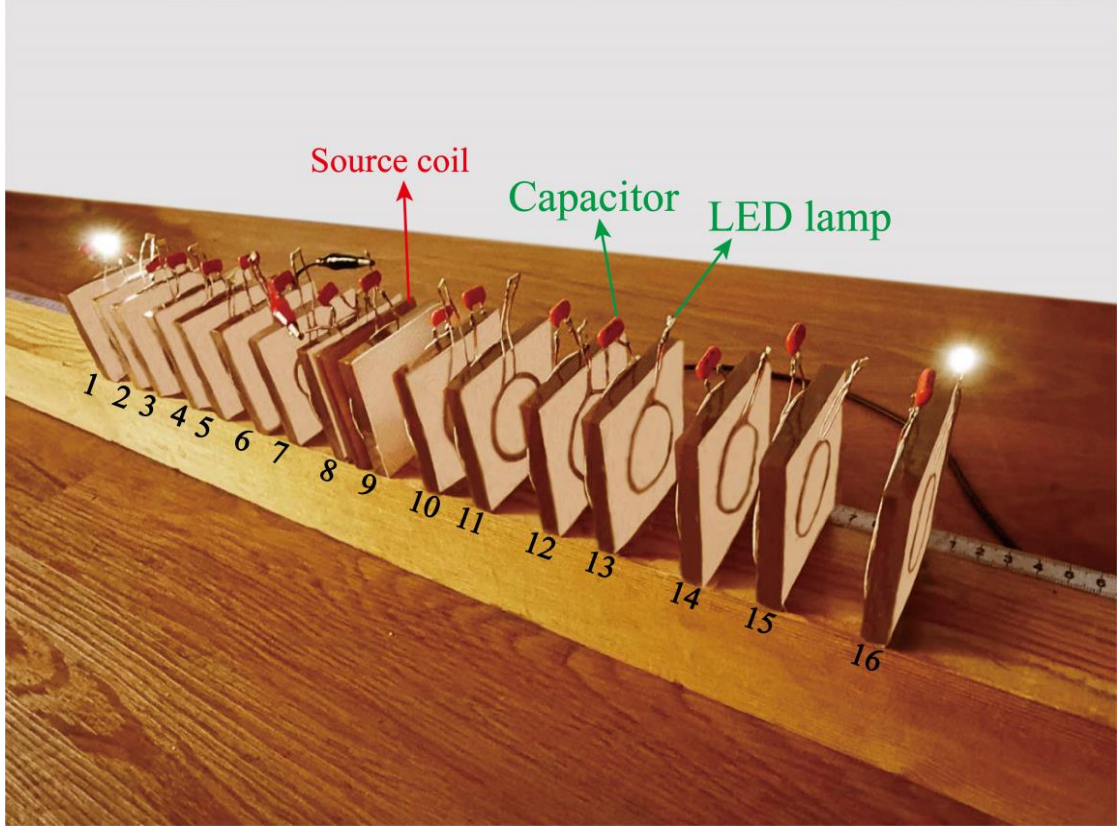

FIG. 5. **Experimental demonstration of the 1-W WPT for lighting two LED lamps.** The non-resonant source coil is placed in the center of the chain. All resonators are added with LED lamps except for the two resonators near the source coil. At the working frequency 5.62 MHz, the TEM can obviously light up the LED lamps on both sides of the chain, while the LED lamps loaded on other composite resonators remain dark.

## IV. AN EFFECTIVE THIRD-ORDER *PT* SYMMETRIC SYSTEM IMPLEMENTED BY TOPOLOGICAL MODES FOR WPT

In the above section, we compared the influence of two topological distinguished configurations of dimer chains on WPT in detail. We found that topological dimer chain has a significant enhancement effect on the long-range WPT than trivial chain. However, there are some aspects of topological dimer chain for long-range WPT that should be noted. On the one hand, for the topological dimer chain with second-order *PT* symmetry, the standby power loss is inevitable. On the other hand, the TEMs at the two ends of the structure will interact and lead to frequency splitting as a result of near-field coupling [20, 63]. This effect not only weakens the robustness of the TEM, but also brings the problem of frequency tracking at the purely real eigenfrequency in topological WPT [55, 56]. Recently, stable and efficient power transfer in three-coil WPT systems have been studied, in which a coupling independent real eigenfrequency can be used to stable WPT without frequency tracking [59, 60]. Here, an effective third-order *PT* symmetry, based on the near-field coupling between three topological modes in a composite chain composed of two topological dimer chains, is proposed to overcome above mentioned limitations. Specially, the effective third-order *PT* symmetry is realized by two TEMs and one TIM, as schematically shown in Fig. 6.

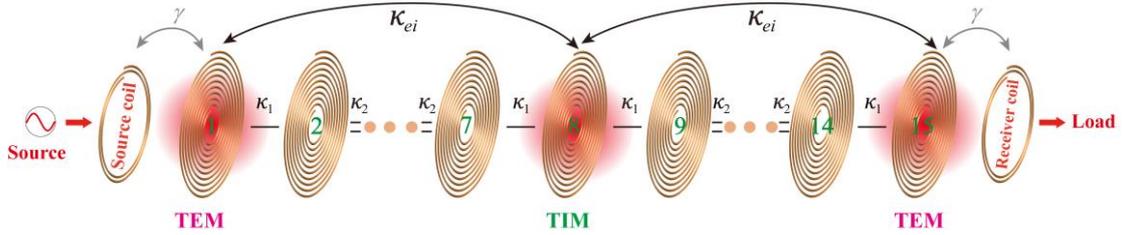

FIG. 6. **Schematic of a multi-coil WPT system based on the third-order *PT* symmetry in a composite topological dimer chain ($\kappa_1 < \kappa_2$).** The effective third-order *PT* symmetry is formed by the interaction of the three topological modes, including two TEMs at two ends of the chain and one TIM at the center of the chain.

Similar to Eqs. (2) and (3), the theoretical analysis of TEMs in the effective topological chain with third-order *PT* symmetry is realized based on the coupled mode equations and effective Hamiltonian. Here, $\kappa_1 = 0.55$ MHz ($d_1 = 2.5$ cm) and $\kappa_2 = 0.63$ MHz ($d_2 = 2$ cm). There are three topological modes induced by the near-field coupling between two TEMs and one TIM in the bandgap, as shown in Fig. 7(a). The frequencies of two topological modes deviated from the working frequency while one topological mode is fixed, similar to the simple three-coil WPT systems [59, 60]. Specially, the normalized intensity distributions of three topological modes are shown in Figs. 7(b)-7(d), respectively. The second topological mode is fixed at the working frequency, which can be used to realize the stable long-range WPT. Specially, figure 7(c) shows that the field amplitude of the second topological mode is symmetrically localized at the two ends of the chain while the inner intensity of the chain is low.

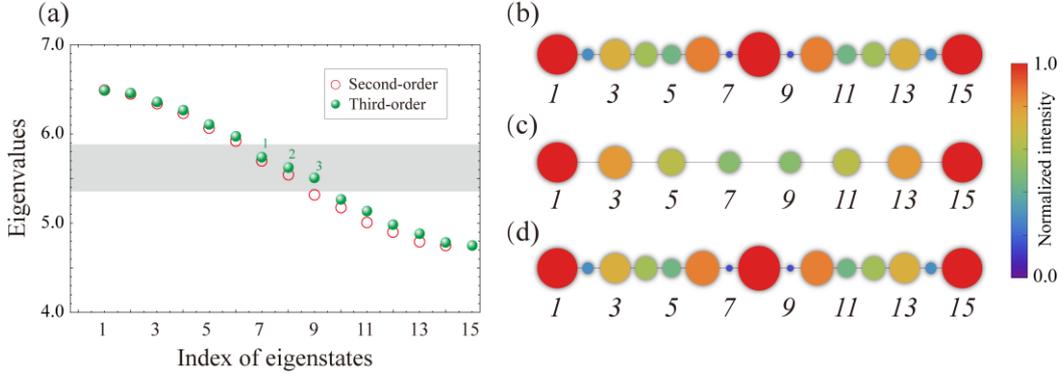

FIG. 7. **Three coupled topological modes in the composite topological dimer chain.** (a) The eigenvalues spectra of the topological dimer chains with third-order $PT$ symmetry (green dots) and second-order $PT$ symmetry (red circles). Discrete topological modes are identified at the bandgap (marked by the gray region) of the chain. (b)-(d) Normalized intensity distributions of three topological modes in the composite topological dimer chain with third-order $PT$ symmetry.

In practice, we expect that the WPT devices to be valid for diversified loads with different power. We first study the real and imaginary parts of the eigenfrequencies of the effective second-order $PT$ symmetric system implemented by TEMs in topological dimer chains. Specially, the evolution of the real part of eigenfrequencies is calculated in Fig. 8(a). The TEMs coalesce at an $EP_2$ with increasing the power of the loads. It should be noted that the power of the loads is represented by the coupling rate between the TEM on the end of the dimer chain and the receiver. After undergoing the $EP_2$, their imaginary parts are repelled, which results in an increasing imaginary part for one of the eigenfrequencies and a decreasing imaginary part for the other. Therefore, it can be clearly seen that the WPT system with effective second-order *PT* symmetry can have high efficiency in the strongly coupled region, by tuning the working frequency to track the load-dependent real eigenfrequency. Once the power of the load deviates from the optimal value, the transfer efficiency will decrease rapidly. In contrast, the effective Hamiltonian of the third-order open system can be expressed as [60-62]:

$$H_{eie} = \begin{pmatrix} \omega_0 + i\gamma & \kappa_{ei} & 0 \\ \kappa_{ei} & \omega_0 & \kappa_{ei} \\ 0 & \kappa_{ei} & \omega_0 - i\gamma \end{pmatrix}, \qquad (4)$$

where $\kappa_{ei}$ denotes the effective coupling coefficient of the TEM and TIM. One of the eigenfrequencies (5.62 MHz) of the topological WPT system with effective third-order *PT* symmetry is real independent of the load power, as shown in Figs. 8(c) and 8(d). From this point of view, such high-order *PT* symmetry, with load-independent real eigenfrequency, can be regarded as a highly efficient and practical WPT strategy without frequency tracking.

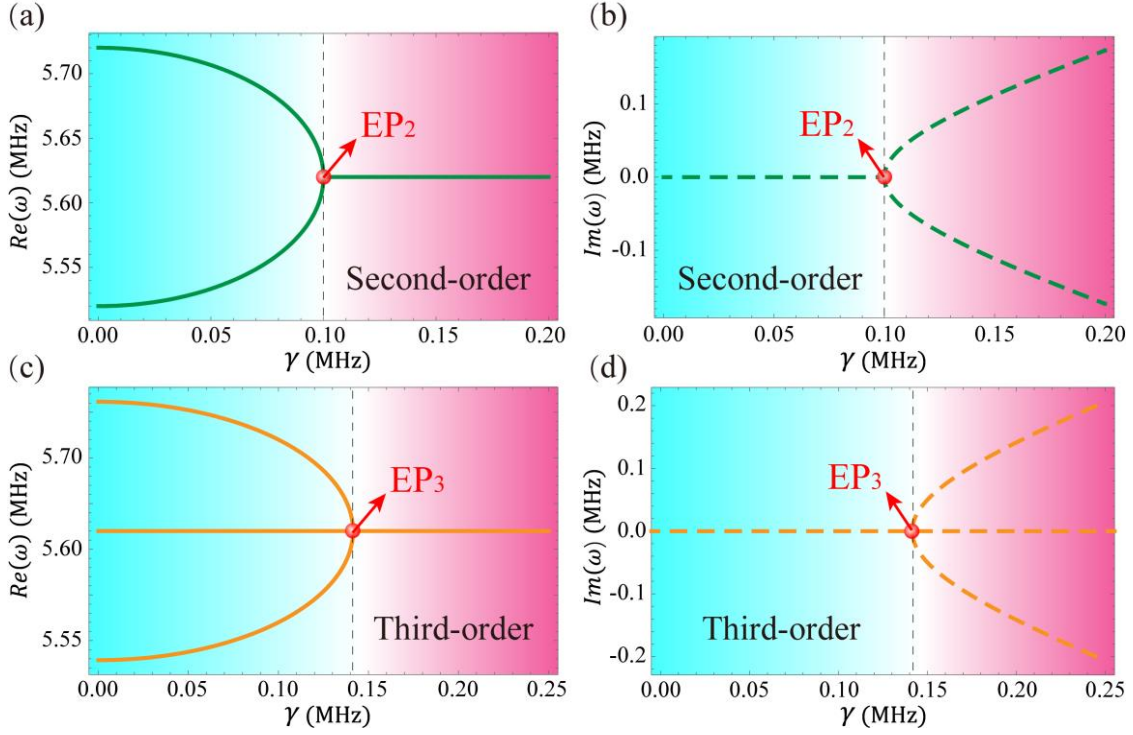

FIG. 8. **Calculated evolution of real part and imaginary part of eigenfrequencies with different power loads.** Real (a) and imaginary (b) eigenfrequencies of topological WPT system with second-order *PT* symmetry. The coupling coefficient between two TEMs is $\kappa_{ee}$ = 0.1 MHz. (c) (d) Similar to (a) (b) but for the topological WPT system with third-order *PT* symmetry. The coupling coefficient between the TEM and TIM is $\kappa_{ei}$ = 0.1 MHz. The frequency of both the TEM and TIM is 5.62 MHz.

Furthermore, we study the transmittance performance of the topological dimer chain with third-order *PT* symmetry. Figure 9(a) shows the comparison of calculated transmission between the topological dimer chains with third-order *PT* symmetry and second-order *PT* symmetry. We can clearly see that there is a transmission peak at the reference frequency (5.62 MHz) in the chain with third-order *PT* symmetry. In this case, the topological mode fixed at the working frequency can be used to realize the stable WPT. The enhancement of transmission efficiency of topological dimer chains with third-order *PT* symmetry compared with second-order *PT* symmetry is shown in Fig. 9(b). The deviation of experimental and theoretical results mainly comes from the loss and construction error of the samples. It should be noted that because the difference between the intra-chain coupling $\kappa_1$ and the inter-chain coupling $\kappa_2$ is reduced, there is an increase in the transmittance of the trivial chain at the working frequency, thus the transmission efficiency enhancement here is not significant enough. Nevertheless, the topological dimer chains with third-order $PT$ symmetry can solve the frequency tracking problem of the topological dimer chains with second-order $PT$ symmetry.

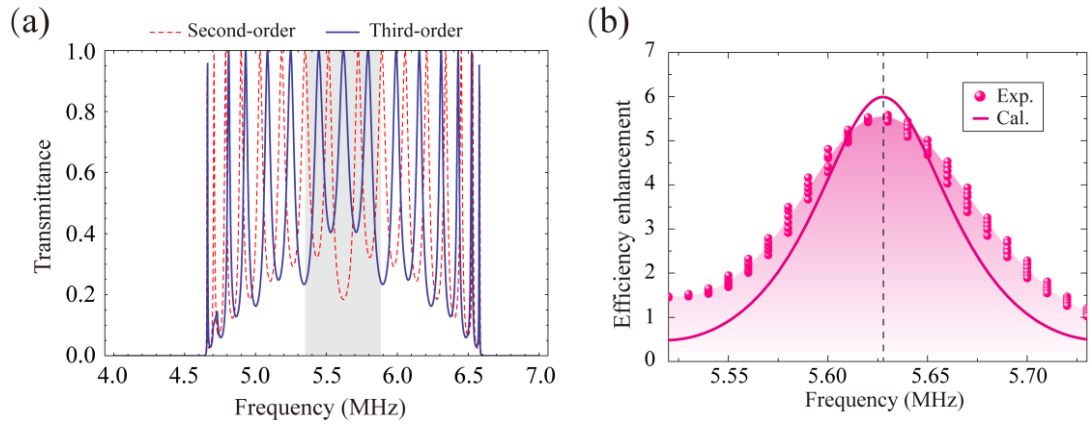

FIG. 9. **Enhancement of transmission efficiency of the topological dimer chain with third-order *PT* symmetry compared with second-order symmetry.** (a) Calculated transmittance spectra of the topological chain with third-order $PT$ symmetry (solid blue line) and second-order $PT$ symmetry (red dashed line). (b) The transmission efficiency of a topological chain with third-order $PT$ symmetry divided by that of a chain with second-order $PT$ symmetry.

At the end of this section, an important technical problem of WPT system is studied, which is standby power loss. In fact, it is very meaningful to keep the low energy output of the system when the system is standby. Figures 10(a) and 10(c) show the measured reflection spectrum of the topological dimer chain with effective third-order $PT$ symmetry under working and standby states, respectively. Specially, in contrast to the working state, the standby state corresponds to the case that one resonator at the right end of the chain is removed. From Figs. 10(a) and 10(c), it can be clearly seen that the refection of the chain under working (standby) state is low (high) at the reference frequency, which means the standby power loss of the topological dimer chain with effective third-order *PT* symmetry is small. Then, the characteristics of low standby power loss are intuitively demonstrated. The source coil is placed at the left end of the chain and both ends of the chain are equipped with an LED lamp. Two LED lamps will be lighted up at the working state, as shown in Fig. 10(b). However, for the standby state, two LED lamps remain dark at the reference frequency in Fig. 10(d). Compared with Figs. 10(b) and 10(d), we further demonstrate that the standby power loss of the topological dimer chain with effective third-order *PT* symmetry is small.

In addition, we evaluate the standby power loss of the topological dimer chain with effective second-order *PT* symmetry in Figs. 10(e)-10(h) for comparison. In Fig. 10(e), the reflection of the topological dimer chain with effective second-order *PT* symmetry under the working state is also low, which is similar to the case of the third-order topological chain in Fig. 10(a). And the LED lamps placed at two ends of the chain can also be lighted up by the left source coil, which is shown in Fig 10(f). However, the refection of the topological dimer chain with effective second-order *PT* symmetry under standby state is low at the reference frequency (Fig. 10(g)), which means the standby power loss of this topological chain is large. The large standby power loss of the topological dimer chain with effective second-order *PT* symmetry is demonstrated in Fig. 10(h). In contrast to Fig. 10(d), the LED lamp at the left end of the chain will be lighted up when the source coil is placed at the left end of the topological dimer chain with effective second-order *PT* symmetry in Fig. 10(h). Therefore, although the topological dimer chain with effective second-order *PT* symmetry can achieve efficient energy transmission, standby power loss is still an important problem for practical application. Compared with Figs. 10(a)-10(d) and Figs. 10(e)-10(h), we can see that the topological structure with effective third-order *PT* symmetry can better solve the standby power loss problem of the topological structure with effective second-order *PT* symmetry. In a word, the topological modes of dimer chains can be used to achieve robust long-range WPT. In particular, for the system with effective third-order *PT* symmetric, it has obvious advantages in practical application, because it can also solve the problems of frequency tracking and standby power loss.

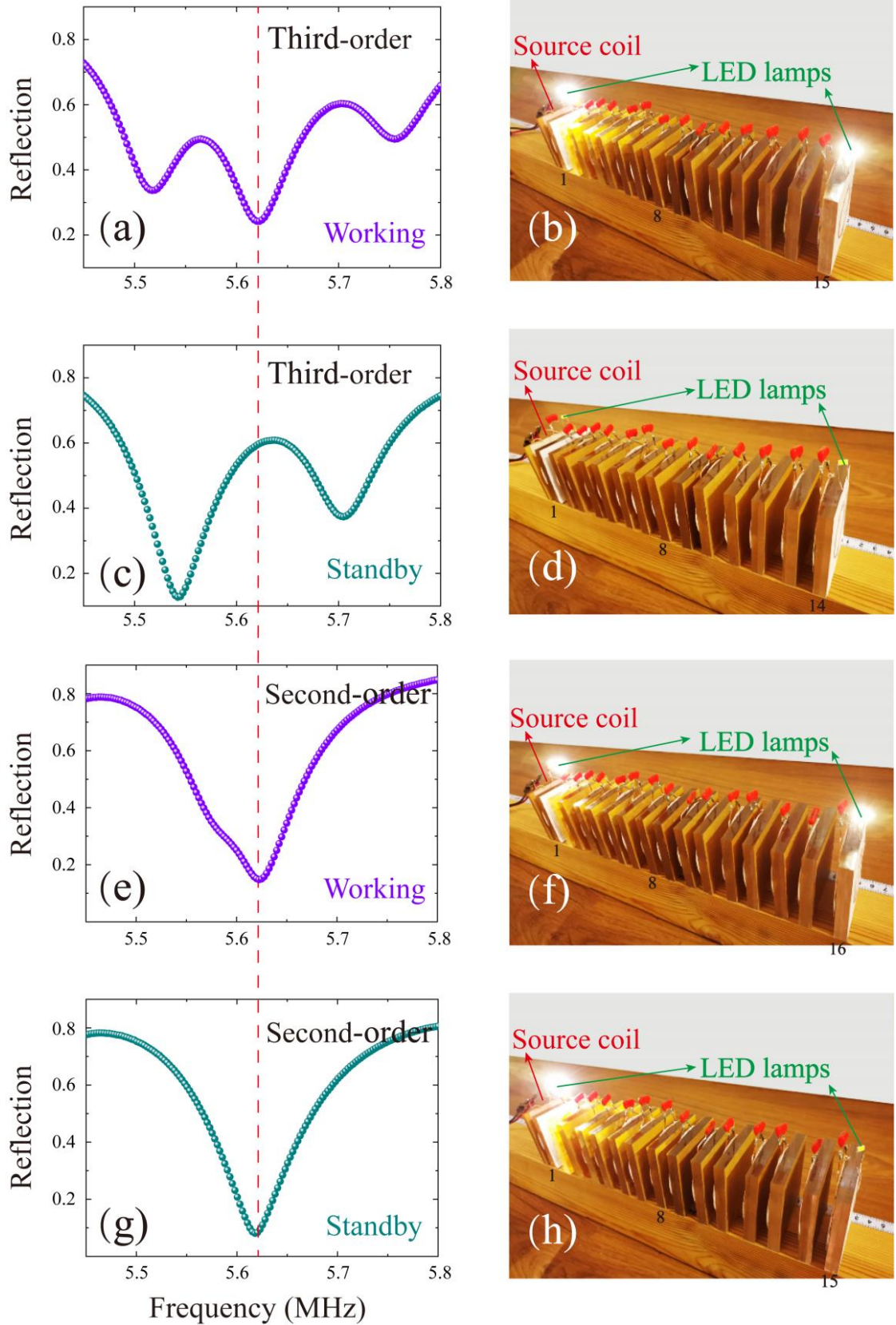

FIG. 10. **Experimental demonstration of the WPT with small standby power loss in the topological dimer chain with third-order *PT* symmetry.** (a) Measured reflection spectrum of the topological dimer chain with effective third-order *PT* symmetry in the working state. (b) Experimental demonstration of the topological WPT of topological dimer chain with effective third-order PT symmetry by lighting two LED lamps at two ends of the chain. (c) Similar to (a), but one resonator at the right end of the chain is removed, which corresponds to the standby state. (d) Experimental demonstration of the standby power loss in the topological dimer chain with effective third-order *PT* symmetry under standby state is small because the LED lamps remain dark. (e)-(h) Similar to (a)-(d), but for the topological dimer chain with effective second-order *PT* symmetry under working and standby states. The LED lamp at the left end of the chain is illuminated even if the system is standby, which means the standby power loss is large.

## V. CONCLUSION

In summary, based on the topological dimer chain composed of ultra-subwavelength coil resonators, we experimentally verify that the TEMs in the second-order *PT* symmetry WPT structure can achieve efficient long-range WPT. This topological WPT inherits the physical properties of TEMs and has good electromagnetic compatibility because it is immune to impurities and perturbations. Specially, in order to solve the technical problems of standby power and frequency tracking, we also propose the third-order *PT* symmetry WPT structure with two TEMs and one TIM. This work provides a versatile platform with topological protection for long-range WPT, which has the potential to explore high-performance WPT devices with more complex structures, including the trimer and quasiperiodic chains. Moreover, our results may be extended to other physical platforms such as sound [70] and heat [71] transfer.

**ACKNOWLEDGMENTS**

This work was supported by the National Key R&D Program of China (Grant No. 2016YFA0301101), the National Natural Science Foundation of China (NSFC; Grant Nos. 12004284, 11774261, and 61621001), the Natural Science Foundation of Shanghai (Grant Nos. 18JC1410900), the China Postdoctoral Science Foundation (Grant Nos. 2019TQ0232 and 2019M661605), and the Shanghai Super Postdoctoral Incentive Program.

**Note added** – Recently, we found an independent investigation of topological WPT using topological dimer chain beyond the traditional domino-form coil chain [72]. However, the problems of the standby power and the frequency tracking have not been studied based on the effective second-order *PT* system in [72].

**References**

1. L. Lu, J. D. Joannopoulos, and M. Soljačić, Topological photonics, *Nat. Photon.* **8**, 821 (2014).

2. Y. Wu, C. Li, X. Y. Hu, Y. T. Ao, Y. F. Zhao and Q. H. Gong, Applications of topological photonics in integrated photonic devices, *Adv. Opt. Mater.* **5**, 1700357 (2017).

3. T. Ozawa, H. M. Price, A. Amo, N. Goldman, M. Hafezi, L. Lu, M. C. Rechtsman, D. Schuster, J. Simon, O. Zilberberg, and I. Carusotto, Topological photonics, *Rev. Mod. Phys.* **91**, 015006 (2019).

4. A. B. Khanikaev, S. H. Mousavi, W. K. Tse, M. Kargarian, A. H. MacDonald, and G. Shvets, Photonic topological insulators, *Nat. Mater.* **12**, 233-239 (2013).

5. Z. W. Guo, H. T. Jiang, Y. Long, K. Yu, J. Ren, C. H. Xue, and H. Chen, Photonic spin Hall effect in waveguides composed of two types of single-negative metamaterials, *Sci. Rep.* **7**, 7742 (2017).

6. Y. Li, Y. Sun, W. W. Zhu, Z. W. Guo, J. Jiang, T. Kariyado, H. Chen, and X. Hu, Topological LC-circuits based on microstrips and observation of electromagnetic modes with orbital angular momentum, *Nat. Commun.* **9**, 4598 (2018).

7. H. Jia, R. Zhang, W. Gao, Q. Guo, B. Yang, J. Hu, Y. Bi, Y. Xiang, C. Liu, S. Zhang, Observation of chiral zero mode in inhomogeneous three dimensional Weyl metamaterials, *Science* **363**, 148 (2019).

8. Y. H. Yang, Z. Gao, H. R. Xue, L. Zhang, M. J. He, Z. J. Yang, R. Singh, Y. D. Chong, B. L. Zhang and H. S. Chen, Realization of a three-dimensional photonic topological insulator, *Nature*

**565**, 622 (2019).

9. M. Hafezi, E. A. Demler, M. D. Lukin, and J. M. Taylor, Robust optica delay lines with topological protection, *Nat. Phys.* **7**, 907-912 (2011).

10. G. Q. Liang and Y. D. Chong, Optical resonator analog of a two-dimensional topological insulator, *Phys. Rev. Lett.* **110**, 203904 (2013).

11. Y. Ao, X. Y. Hu, Y. L. You, C. C. Lu, Y. L. Fu, X. Y. Wang, and Q. H. Gong, Topological phase transition in the non-Hermitian coupled resonator array, *Phys. Rev. Lett.* **125**, 013902 (2020).

12. Z. W. Guo, Y. Sun, H. T. Jiang, Y. Q. Ding, Y. H. Li, Y. W. Zhang, H. Chen, Experimental demonstration of an anomalous Floquet topological insulator based on negative-index media, arXiv: 2006. 12252 (2020).

13. G. Harari, M. A. Bandres, Y. Lumer, M. C. Rechtsman, Y. D. Chong, M. Khajavikhan, D. N. Christodoulides, and M. Segev, Topological insulator laser: Theory, *Science* **359**, eaar4003 (2018).

14. M. A. Bandres, S. Wittek, G. Harari, M. Parto, J. Ren, M. Segev, D. N. Christodoulides, and M. Khajavikhan, Topological insulator laser: Experiments, *Science* **359**, eaar4005 (2018).

15. Z. K. Shao, H. Z. Chen, S. Wang, X. R. Mao, Z. Q. Yang, S. L. Wang, X. X. Wang, X. Hu and R. M. Ma, A high-performance topological bulk laser based on band-inversion-induced reflection, *Nat. Nanotech.* **15**, 67 (2019).

16. Z. Q. Yang, Z. K. Shao, H. Z. Chen, X. R. Mao, and R. M. Ma, Spin-momentum-locked edge mode for topological vortex lasing, *Phys. Rev. Lett.* **125**, 013903 (2020).

17. W. P. Su, J. R. Schrieffer, and A. J. Heeger, Solitons in polyacetylene, *Phys. Rev. Lett.* **42**, 1698 (1979).

18. N. Malkova, I. Hromada, X. Wang, G. Bryant, and Z. Chen, Observation of optical Shockley-

like surface states in photonic superlattices, *Opt. Lett.* **34**, 1633 (2009).

19. H. Liu, D. A. Genov, D. M. Wu, Y. M. Liu, Z. W. Liu, C. Sun, S. N. Zhu, and X. Zhang, Magnetic plasmon hybridization and optical activity at optical frequencies in metallic nanostructures, *Phys. Rev. B* **76**, 073101 (2007).

20. J. Jiang, J. Ren, Z. W. Guo, W. W. Zhu, Y. Long, H. T. Jiang, and H. Chen, Seeing topological winding number and band inversion in photonic dimer chain of split-ring resonators, *Phys. Rev. B* **101**, 165427 (2020).

21. F. Zangeneh-Nejad and R. Fleury, Disorder-induced signal filtering with topological metamaterials, *Adv. Mat.* **32**, 2001034 (2020).

22. Q. S. Huang, Z. W. Guo, J. T. Feng, C. Y. Yu, H. T. Jiang, Z. Zhang, Z. S. Wang, and H. Chen, Observation of a topological edge state in the X-ray band, *Laser & Photon. Rev.* **13**, 1800339 (2019).

23. Y. Hadad, J. C. Soric, A. B. Khanikaev, and A. Alù, Self-induced topological protection in nonlinear circuit arrays, *Nat. Electron.* **1**, 178 (2018).

24. D. Smirnova, D. Leykam, Y. D. Chong, and Y. Kivshar, Nonlinear topological photonics, *Appl. Phys. Rev.* **7**, 021306 (2020).

25. S. Xia, D. Jukić, N. Wang, D. Smirnova, L. Smirnov, L. Tang, D. Song, A. Szameit, D. Leykam, J. Xu, Z. Chen, and H. Buljan, Nontrivial coupling of light into a defect: the interplay of nonlinearity and topology, *Light Sci. Appl.* **9**, 147 (2020).

26. H. Zhao, P. Miao, M. H. Teimourpour, S. Malzard, R. El-Ganainy, H. Schomerus and L. Feng, Topological hybrid silicon microlasers, *Nat. Commun.* **9**, 981 (2018).

27. M. Parto, S. Wittek, H. Hodaei, G. Harari, M. A. Bandres, J. Ren, M. C. Rechtsman, M. Segev, D. N. Christodoulides, and M. Khajavikhan, Edge-mode lasing in 1D topological active arrays, *Phys.*

*Rev. Lett*. **120**, 113901 (2018).

28. C. Poli, M. Bellec, U. Kuhl, F. Mortessagne, and H. Schomerus, Selective enhancement of topologically induced interface states in a dielectric resonator chain, *Nat. Commun.* **6**, 6710 (2015).

29. S. Weimann, M. Kremer, Y. Plotnik, Y. Lumer, S. Nolte, K. G. Makris, M. Segev, M. C. Rechtsman, and A. Szameit, Topologically protected bound states in photonic parity-time-symmetric crystals, *Nat. Mater.* **16**, 433 (2017).

30. B. K. Qi, L. X. Zhang, and Li Ge, Defect states emerging from a non-Hermitian flatband of photonic zero modes, *Phys. Rev. Lett.* **120**, 093901 (2018).

31. S. Longhi, Topological phase transition in non-Hermitian quasicrystals, *Phys. Rev. Lett.* **122**, 237601 (2019).

32. W. Song, W. Z. Sun, C. Chen, Q. H. Song, S. M. Xiao, S. N. Zhu, and T. Li, Breakup and recovery of topological zero modes in finite non-Hermitian optical lattices, *Phys. Rev. Lett.* **123**, 165701 (2019).

33. Z. W. Guo, T. Z. Zhang, J. Song, H. T. Jiang, and H. Chen, Sensitivity of topological edge states in a non-Hermitian dimer chain, arXiv: 2007.04409 (2020).

34. A. Kurs, A. Karalis, R. Moffatt, J. D. Joannopoulos, P. Fisher, and M. Soljačić, Wireless power transfer via strongly coupled magnetic resonances, *Science* **317**, 83 (2007).

35. A. P. Sample, D. A. Meyer, and J. R. Smith, Analysis, experimental results, and range adaptation of magnetically coupled resonators for wireless power transfer, *IEEE Trans. Ind. Electron.* **58**, 544 (2011).

36. S. Kim, J. S. Ho, and A. S. Poon, Midfield Wireless Powering of subwavelength autonomous devices, *Phys. Rev. Lett.* **110**, 203905 (2013).

37. M. Song, P. Belov, and P. Kapitanova, Wireless power transfer inspired by the modern trends in electromagnetics, *Appl. Phys. Rev.* **4**, 021102 (2017).

38. K. Sun, R. Fan, X. Zhang, Z. Zhang, Z. Shi, N. Wang, P. Xie, Z. Wang, G. Fan, H. Liu, C. Liu, T. Li, C. Yan, and Z. Guo, An overview of metamaterials and their achievements in wireless power transfer, *J. Mater. Chem. C* **6**, 2925 (2018).

39. S. Y. R. Hui and W. C. Ho, A new generation of universal contactless battery charging platform for portable consumer electronic equipment, *IEEE Trans. Power Electron.* **20**, 620 (2005).

40. Y. Urzhumov and D. R. Smith, Metamaterial-enhanced coupling between magnetic dipoles for efficient wireless power transfer, *Phys. Rev. B* **83**, 205114 (2011).

41. Q. Wu, Y. H. Li, N. Gao, F. Yang, Y. Q. Chen, K. Fang, Y. W. Zhang, and H. Chen, Wireless power transfer based on magnetic metamaterials consisting of assembled ultra-subwavelength meta-atoms, *Europhys. Lett.* **109**, 68005 (2015).

42. M. Z. Song, K. Baryshnikova, A. Markvart, P. Belov, E. Nenasheva, C. Simovski, and P. Kapitanova, Smart table based on a metasurface for wireless power transfer, *Phys. Rev. Appl.* **11**, 054046 (2019).

43. Y. Q. Chen, Z. W. Guo, Y. Q. Wang, X. Chen, H. T. Jiang, and H. Chen, Experimental demonstration of the magnetic field concentration effect in circuit-based magnetic near-zero index media, *Opt. Express* **28**, 17064 (2020).

44. B. L. Cannon, J. F. Hoburg, D. D. Stancil, and S. C. Goldstein, Magnetic resonant coupling as a potential means for wireless power transfer to multiple small receivers, *IEEE Trans. Power Electron.* **24**, 1819 (2009).

45. A. P. Sample, D. A. Meyer, and J. R. Smith, Analysis, experimental results, and range adaptation of magnetically coupled resonators for wireless power transfer, *IEEE Trans. Ind. Electron.* **58**, 544 (2011).

46. B. N. Wang, K. H. Teo, T. Nishino, W. Yerazunis, J. Barnwell, and J. Y. Zhang, Experiments on wireless power transfer with metamaterials, *Appl. Phys. Lett.* **98**, 254101 (2011).

47. K. JinWook, S. Hyeon-Chang, K. Kwan-Ho, and P. Young-Jin, Efficiency analysis of magnetic resonance wireless power transfer with intermediate resonant coil, *IEEE Antennas Wireless Propag. Lett.* **10**, 389 (2011).

48. J. S. Hoa, A. J. Yeha, E. Neofytoub, S. Kima, Y. Tanabea, B. Patlollab, R. E. Beyguib, and A. S. Y. Poon, Wireless power transfer to deep-tissue microimplants, *Proc. Natl Acad. Sci. USA* **111**, 7974 (2014).

49. H. Mei and P. P. Irazoqui, Miniaturizing wireless implants, *Nat. Biotechnol.* **32**, 1008 (2014).

50. H. Li, J. Li, K. Wang, W. Chen, and X. Yang, A maximum efficiency point tracking control scheme for wireless power transfer systems using magnetic resonant coupling, *IEEE Trans. Power Electron.* **30**, 3998 (2015).

51. A. Krasnok, D. G. Baranov, A. Generalov, S. Li, and A. Alù, Coherently enhanced wireless power transfer, *Phys. Rev. Lett.* **120**, 143901 (2018).

52. C. Saha, I. Anya, C. Alexandru, and R. Jinks, Wireless power transfer using relay resonators, *Appl. Phys. Lett.* **112**, 263902 (2018).

53. C. Badowich and L. Markley, Idle power loss suppression in magnetic resonance coupling wireless power transfer, *IEEE Trans. Ind. Electron.* **65**, 8605 (2018).

54. J. L. Zhou, B. Zhang, W. X. Xiao, D. Y. Qiu, and Y. F. Chen, Nonlinear parity-time-symmetric

model for constant efficiency wireless power transfer: Application to a drone-in-flight wireless charging platform, *IEEE Trans. Ind. Electron.* **66**, 4097 (2019).

55. S. Assawaworrarit, X. F Yu, and S. H. Fan, Robust wireless power transfer using a nonlinear parity–time-symmetric circuit, *Nature* **546**, 387 (2017).

56. G. Lerosey, Applied physics: Wireless power on the move, *Nature* **546**, 354 (2017).

57. S. Assawaworrarit and S. H. Fan, Robust and efficient wireless power transfer using a switch-mode implementation of a nonlinear parity–time symmetric circuit, *Nat. Electron.* **3**, 273 (2020).

58. H. Li, J. Li, K. Wang, W. Chen, and X. Yang, A maximum efficiency point tracking control scheme for wireless power transfer systems using magnetic resonant coupling, *IEEE Trans. Power Electron.* **30**, 3998 (2015).

59. M. Sakhdari, M. Hajizadegan, and P. Y. Chen, Robust extended-range wireless power transfer using a higher-order PT-symmetric platform, *Phys. Rev. Res.* **2**, 013152 (2020).

60. C. Zeng, Y. Sun, G. Li, Y. H. Li, H. T. Jiang, Y. P. Yang, and H. Chen, High-order parity-time symmetric model for stable three-coil wireless power transfer, *Phys. Rev. Appl.* **13**, 034054 (2020).

61. Y. Sun, W. Tan, H. Q. Li, J. Li, and H. Chen, Experimental demonstration of a coherent perfect absorber with PT phase transition, *Phys. Rev. Lett.* **112**, 143903 (2014).

62. C. Zeng, Y. Sun, G. Li, Y. Li, H. Jiang, Y. P. Yang and H. Chen, Enhanced sensitivity at high-order exceptional points in a passive wireless sensing system, *Opt. Express* **27**, 27562 (2019).

63. J. Jiang, Z. W. Guo, Y. Q. Ding, Y. Sun, Y. H. Li, H. T. Jiang, and H. Chen, Experimental demonstration of the robust edge states in a split-ring-resonator chain, *Opt. Express* **26**, 12891 (2018).

64. W. X. Zhong, C. K. Lee, and S. Y. Ron Hui, General analysis on the use of Tesla's resonators

in domino forms for wireless power transfer, *IEEE Trans. Ind. Electron.* **60**, 261 (2013).

65. Z. W. Guo, H. T. Jiang, Y. Sun, Y. H. Li, and H. Chen, Asymmetric topological edge states in a quasiperiodic Harper chain composed of split-ring resonators, *Opt. Lett.* **43**, 5142 (2018).

66. X. L. Liu and G. S. Agarwal, The new phases due to symmetry protected piecewise berry phases; enhanced pumping and nonreciprocity in trimer lattices, *Sci. Rep.* **7**, 45015 (2017).

67. Z. W. Guo, H. T. Jiang, Y, H, Li, H. Chen, and G. S. Agarwal, Enhancement of electromagnetically induced transparency in metamaterials using long range coupling mediated by a hyperbolic material, *Opt. Express* **26**, 627 (2018).

68. J. Song, F. Q. Yang, Z. W. Guo, Y. Q. Chen, H. T. Jiang, Y. H. Li, and H. Chen, One-dimensional topological quasiperiodic chain for directional wireless power transfer, arXiv: 2008. 10352 (2020).

69. J. Feis, C. J. Stevens, and E. Shamonina, Wireless power transfer through asymmetric topological edge states in diatomic chains of coupled meta-atoms, *Appl. Phys. Lett.* **117**, 134106 (2020).

70. Y. Long and J. Ren, Floquet topological acoustic resonators and acoustic Thouless pumping, *J Acoust. Soc. Am.* **146**, 742 (2019).

71. G. M. Tang, H. H. Yap, J. Ren, and J. S. Wang, Anomalous near-field heat transfer in carbon-based nanostructures with edge states, *Phys. Rev. Appl.* **11**, 031004 (2019).

72. L. Zhang, Y. H. Yang, Z. Jiang, Q. L. Chen, Q. H. Yan, Z. Y. Wu, B. L. Zhang, J. T. Huangfu, and H. S. Chen, Topological wireless power transfer, arXiv: 2008. 02592 (2020).